%% file: icme2025_template_anonymized.tex
\documentclass[conference]{IEEEtran}
\IEEEoverridecommandlockouts
\usepackage{cite}
\usepackage{amsmath,amssymb,amsfonts}
\usepackage{algorithmic}
\usepackage{graphicx}
\usepackage{textcomp}
\usepackage{xcolor} 
\usepackage{multirow}
\usepackage{subcaption} 
\usepackage{amsmath}
\usepackage{booktabs} 
\usepackage{float} 
\usepackage{diagbox} 
\usepackage{hyperref}

\def\BibTeX{{\rm B\kern-.05em{\sc i\kern-.025em b}\kern-.08em
    T\kern-.1667em\lower.7ex\hbox{E}\kern-.125emX}}
\begin{document}

\title{Visual-based spatial audio generation system for multi-speaker environments}

\author{
    Xiaojing Liu$^{1,*}$, Ogulcan Gurelli$^{1}$, Yan Wang$^{2,*}$, and Joshua Reiss$^{1}$ \\
    $^{1}$Queen Mary University of London, UK \\
    $^{2}$Xidian University, China \\
    Emails: \texttt{\{xiaojing.liu, ec21698, joshua.reiss\}@qmul.ac.uk} \\ 
    \texttt{wangyan97@stu.xidian.edu.cn} \\
    $^{*}$Corresponding authors
}

\maketitle

\begin{abstract}
In multimedia applications such as films and video games, spatial audio techniques are widely employed to enhance user experiences by simulating 3D sound: transforming mono audio into binaural formats. However, this process is often complex and labor-intensive for sound designers, requiring precise synchronization of audio with the spatial positions of visual components. To address these challenges, we propose a visual-based spatial audio generation system - an automated system that integrates face detection YOLOv8 for object detection, monocular depth estimation, and spatial audio techniques. Notably, the system operates without requiring additional binaural dataset training. The proposed system is evaluated against existing Spatial Audio generation system using objective metrics. Experimental results demonstrate that our method significantly improves spatial consistency between audio and video, enhances speech quality, and performs robustly in multi-speaker scenarios. By streamlining the audio-visual alignment process, the proposed system enables sound engineers to achieve high-quality results efficiently, making it a valuable tool for professionals in multimedia production.
\end{abstract}

\begin{IEEEkeywords}
Audio-Visual Synchronization, Spatial Audio, Speech Quality, Multi-Speaker, Post-Production.
\end{IEEEkeywords}

\section{Introduction}
\label{sec:intro}

In multimedia production, including films, advertisements, teleconferencing and video games, achieving seamless alignment between character dialogue and corresponding visual elements is essential for creating immersive experiences. As humans rely on multi-modal cues to interpret and engage with real-world events \cite{10687782}, the demand for high-quality audio-visual experiences continues to grow. With the rise of three dimensional (3D) audio, virtual reality (VR), and augmented reality (AR), the importance of accurate spatial audio alignment has become increasingly evident \cite{8486455}. 
Visual-based audio spatialization has become a prominent area of research due to its broad applications in AR \cite{360audio}, VR \cite{VirtualReality}, social video sharing \cite{4607471}\cite {6012065} and audio-visual video {understanding \cite{AudioVisualUnderstanding}.
Effective audio-visual spatialization enhances realism, enabling audiences to feel as though they are present within the environment. 

Currently, most post-production professionals and audio engineers manually adjust spatial audio parameters on digital platforms, relying on visual cues to determine the positions of sound sources. This process is highly labor-intensive and requires significant time and effort.

Some researchers have already explored generating spatial audio based on video input. Lin et al. \cite{AudioVisualUnderstanding} proposed a model for generating binaural audio from visual frames and monaural audio inputs, demonstrating its effectiveness through comparisons with other models on the FAIR-Play dataset (binaural audio clips recorded in a controlled music room) and the MUSIC-Stereo dataset (a diverse collection of audio-visual clips from musical performances). Ruohan Gao and Kristen Grauman \cite{VisualSound} introduced Mono2 Binaural, a deep network that takes a mixed monaural audio and its accompanying visual frame as input, using a ResNet-18 network to extract visual features and U-NET to extract audio features.  
 
However, these systems rely heavily on large binaural datasets, which pose significant challenges, such as requiring specialized equipment, controlled recording environments, precise audio-visual synchronization, and labor-intensive annotation processes. Additionally, they face issues with overfitting when handling multiple audio tracks, further complicating the training and optimization process. They also struggle to achieve precise spatial positioning with more than two audio sources, frequently leading to compromised sound quality or reduced immersion, ultimately impacting the user experience. 
This lack of fidelity not only affects the user experience but also poses challenges in professional multimedia production environments, where rapid adaptation and precision are essential. 
\begin{figure}[H]
    \centering
    \includegraphics[width=0.45\textwidth, trim={50pt 0pt 378pt 408pt}, clip]{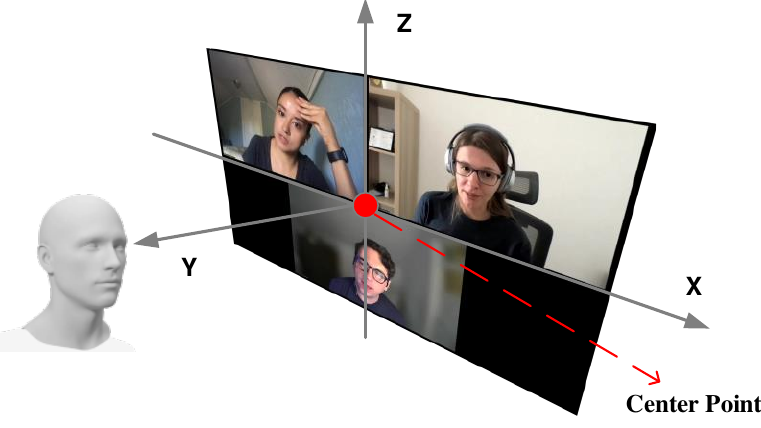}
 % Adjust width as needed
    \caption{Spatialisation of audio representing \textbf{X}, \textbf{Y} and \textbf{Z} coordinates.}
    \label{fig:coordinate}
\end{figure} 

 \begin{figure*}[ht]
    \centering\includegraphics[width=1.0\textwidth, trim={10 149 135 500}, clip]{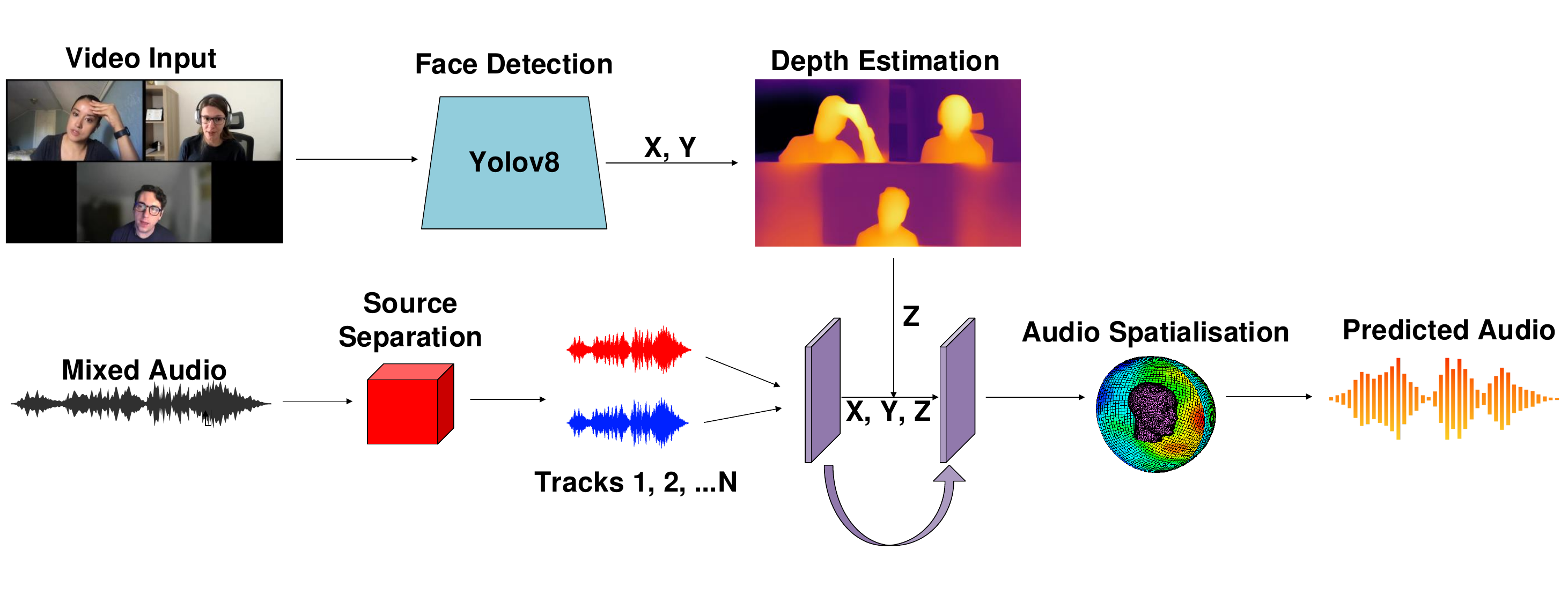} % Adjust trim values as needed
    \caption{Flowgraph of the system showcasing the main processing pipeline.}
    \label{fig:flowgraph}
\end{figure*}

Pseudo Binaural \cite{xu2021visually} has been introduced as an innovative model to generate visually coherent binaural audios from multiple sound sources without requiring recorded binaural data. However, for these system the coordinates for Azimuth and Elevation must be manually pre-defined.

To address these challenges, we propose a visual-based spatial audio generation system designed to support multi-speaker scenarios while eliminating the dependency on large-scale binaural datasets. In this approach (see Fig. \ref{fig:coordinate}), the center of each video frame is assumed to be the origin of a Cartesian coordinate system. Using this framework, the facial positions of individual speakers are accurately calculated, providing spatial cues for generating realistic spatial audio.

Our system integrates object detection using YOLOv8 \cite{sohan2024review} with the WIDER FACE Dataset \cite{yang2016wider} and Depth Estimation model- Depth Anything \cite{depthanything}, to enhance the precision of spatial audio alignment by extracting visual cues with high accuracy, thus facilitating the automatic adjustment of audio sources based on spatial positioning in real-time.

To assess the effectiveness of our system objectively, we compare its performance against existing audio-visual spatial audio generators using established metrics: Perceptual Evaluation of Speech Quality (PESQ), Short-Time Fournier Transform (STFT) Distance, Envelope (ENV) Distance, and Mean Opinion Score (MOS). These metrics enable a comprehensive evaluation of our system’s ability to maintain audio quality and synchronization under varying conditions, demonstrating its applicability in high-fidelity audio post-production environments.

\section{Methodology}

Our model\footnote{\url{https://youtu.be/lWbx58GmP-o}} consists of two main components: visual processing and audio processing. For visual processing, we compare current YOLO models and employ YOLOv8 in combination with the Depth Estimation model Depth Anything (see Sec. \ref{sec:visual}). The details of audio source separation are introduced in Sec. \ref{sec:sourceseparation}. In the audio processing pipeline, we implement two spatialization methods: HRTF convolution and a 3D algorithmic approach, as detailed in Sec. \ref{sectionaudio}. The overall workflow of the system is illustrated in Fig. \ref{fig:flowgraph}.

\subsection{Visual} \label{sec:visual}
Object detection facilitates the localization of spatial positions, particularly azimuthal information, of characters or objects in real-time, acting as anchors for their corresponding audio sources. By generating precise bounding boxes around detected objects, the object detection model ensures consistent spatial alignment, even in dynamic scenes where sources may move or overlap.

YOLO (You Only Look Once) is a widely used object detection model renowned for its speed and accuracy. First introduced by Joseph Redmon et al \cite{redmon2016you} in 2016, YOLO has undergone numerous advancements, with the latest iteration, YOLOv10, incorporating state-of-the-art techniques for enhanced performance and efficiency. This makes it well-suited for applications requiring real-time tracking and localization in complex environments. To evaluate advancements in YOLO's performance and its suitability for real-time applications, we conducted comparative testing across different YOLO versions using the WIDER FACE dataset. This dataset comprises 32,203 images and 393,703 labeled faces, with significant variations in scale, pose, and occlusion. It is divided into training (40\%), validation (10\%), and testing (50\%) sets, ensuring a balanced distribution. The results are summarized in Table~\ref{tab:cap} and Table~\ref{tab:yolo_performance_compact}.

The experiments are conducted using PyTorch 1.7.0 on an NVIDIA RTX 3090 GPU with CUDA 11.0. The system is configured with a 15-core Intel Xeon Platinum CPU and 80 GB RAM, providing sufficient computational resources for training and evaluation.

\vspace{-0.3cm}
\begin{table}[H]
  \centering
  \caption{Performance comparison of YOLO models (P: Precision, R: Recall, mAP50: mean average precision at IoU threshold 0.5, mAP50-95: mean average precision at IoU thresholds ranging from 0.5 to 0.95).}
  \label{tab:cap} 
  \scalebox{1.2}{ 
  \begin{tabular}{lcccc}
    \toprule
    Module       & P     & R     & mAP50 & mAP50-95 \\
    \midrule
    YOLOv5-n     & 0.838 & 0.598 & 0.726 & 0.342    \\
    YOLOv10-n    & 0.829 & 0.554 & 0.633 & 0.316    \\
    YOLOv8-n     & 0.845 & 0.588 & 0.669 & 0.366    \\
    YOLOv-Face2     & \textbf{0.896} & \textbf{0.666} & \textbf{0.735} & \textbf{0.397}    \\
    YOLOv5-s     & 0.872 & 0.655 & 0.696 & 0.347    \\
    \bottomrule
  \end{tabular}}
\end{table} 

\begin{table}[h]
\centering
\caption{Speed comparison of YOLO models (measured in milliseconds, ms).}
\label{tab:yolo_performance_compact}
\scalebox{1.2}{ 
\begin{tabular}{lcccc}
 \toprule
Model & Pre-process & Inference & NMS \\    \midrule
YOLOv5-n       & 0.2                  & 6.1                & 0.8          \\
YOLOv10-n      & 0.1                  & 1.2                & \textbf{0.0}          \\
YOLOv8-n       & \textbf{0.1}                  &\textbf{ 0.6}                & 0.5          \\
YOLOv-Face2    & 16.3        & 1.0        & 17.4 \\
YOLOv5-s       & 16.3                 & 1.0                & 17.4         \\  \toprule
\end{tabular}}
\end{table}

Through Table \ref{tab:cap} and Table \ref{tab:yolo_performance_compact}, bold numbers denote  highest performance. YOLOv8-n \cite{sohan2024review} demonstrates a well-balanced trade-off between accuracy and speed. It achieves a mean average precision of 0.669 (mAP50) and 0.366 (mAP50-95), outperforming YOLOv10-n in both precision (P) and recall (R) metrics while maintaining competitive performance against YOLOv-Face2. Although YOLOv-Face2 achieves slightly higher accuracy, its increased inference time makes it less ideal for real-time applications. YOLOv8-n, with its processing and inference times of just 0.1 ms and 0.6 ms per image, significantly surpasses models like YOLOv5-s and YOLOv-Face2 in speed. This balance of accuracy and efficiency makes YOLOv8-n particularly suitable for scenarios requiring rapid and precise visual tracking to enhance spatial audio generation.

\begin{figure}[H]
    \centering
    \includegraphics[width=0.45\textwidth]{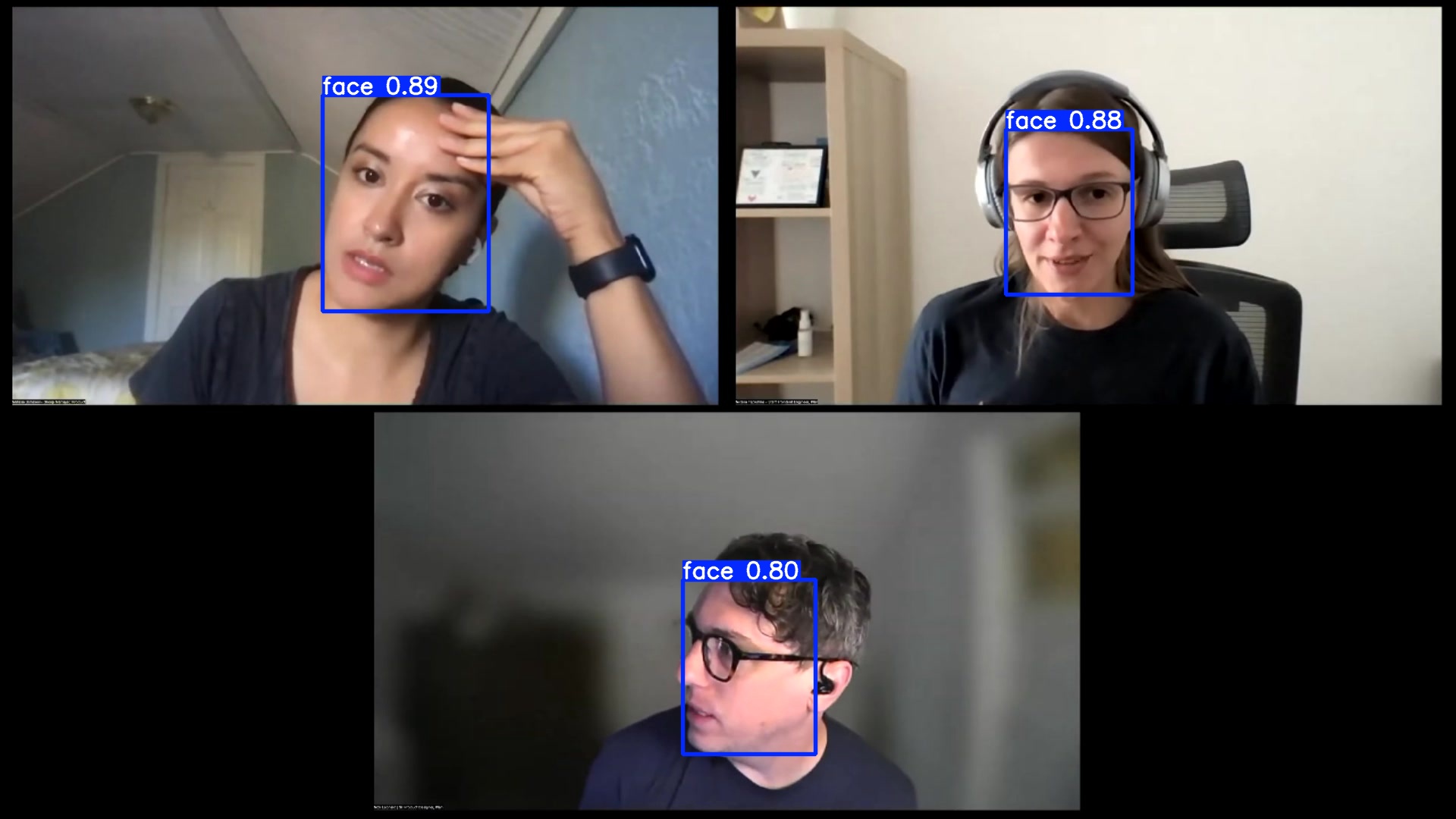} % Adjust width as needed
    \caption{YOLOv8-n Output: Object Detection with Bounding Box Predictions}
    \label{fig:yolov8-output}
\end{figure}

In the results of YOLOv8-n, we obtain the center coordinates \((x_{\text{center}}, y_{\text{center}})\) of the detected bounding box, which are normalized by default. To adapt these coordinates for audio processing, we first normalize \((x_{\text{center}}, y_{\text{center}})\) them into the range \([-1, 1]\).

\begin{equation}
\begin{aligned}
 \textbf{X}&= 2x_{\text{center}} - 1,\\\textbf{Y}&= 1 - 2y_{\text{center}} 
\label{xy}
\end{aligned}
\end{equation}

In three-dimensional space and spatial audio processing, the Cartesian Coordinate System, defined by \textbf{X} (left and right),\textbf{Y} (up and down), and \textbf{Z} (front and back) coordinates. Object detection models, such as YOLOv8-n, provide accurate \textbf{X} and \textbf{Y} coordinates in a 2D plane, enabling real-time detection and localization of individual speakers within an image frame. To extend this to 3D space, depth information (\textbf{Z}-coordinate) must be incorporated, offering a more realistic representation of the environment.

For this purpose, we employ depth estimation techniques to compute the \textbf{Z}-coordinate, representing the distance between the camera and each detected facial individual. Following the methodology outlined in \cite{depthanything}, we utilize the pre-trained ViT-S model from the Depth Anything framework for robust monocular depth estimation. This model excels in handling diverse environments and arbitrary images, delivering precise depth measurements. We chose the Depth Anything ViT-S model for its demonstrated speed efficiency and accuracy in depth estimation tasks, which are crucial for accurately estimating the distance between the sound source and the camera (audience). An example of the results obtained using this model is illustrated in Fig. \ref{fig:deep_estimation}.

In ~\eqref{deepestimation}, $\text{g}_{\text{min}}$ and $\text{g}_{\text{max}}$ represents the range of gray scale value in the depth map, ranging from 0 to 255 \cite{depthanything}. The $\text{O}$ represents the gray scale value calculated from the depth estimation at the location of the image corresponding to each facial detection in YOLOv8-n.
\(d_{\text{min}}\) and \(d_{\text{max}}\) are the minimum and maximum self-defined values of the actual distance  \textbf{Z} in the scene (in meters), set to range from 0.1 to 5.

\begin{equation}
\text{\textbf{Z}} = d_{\text{max}} - (\text{O} - \text{g}_{\text{min}}) \cdot \frac{d_{\text{max}} - d_{\text{min}}}{\text{g}_{\text{max}} - \text{g}_{\text{min}}} 
\label{deepestimation}
\end{equation}

By combining YOLOv8-n's \textbf{X} and \textbf{Y} coordinate with the output \text{\textbf{Z}} calculated from depth estimation model, our approach ensures a higher level of realism and accuracy in tracking and mapping the individual speakers’ position in 3D space.
\begin{figure}[ht] 

    \centering
    \begin{subfigure}[b]{0.241\textwidth} % 宽度改为 0.4
        \centering
        \includegraphics[width=\textwidth]{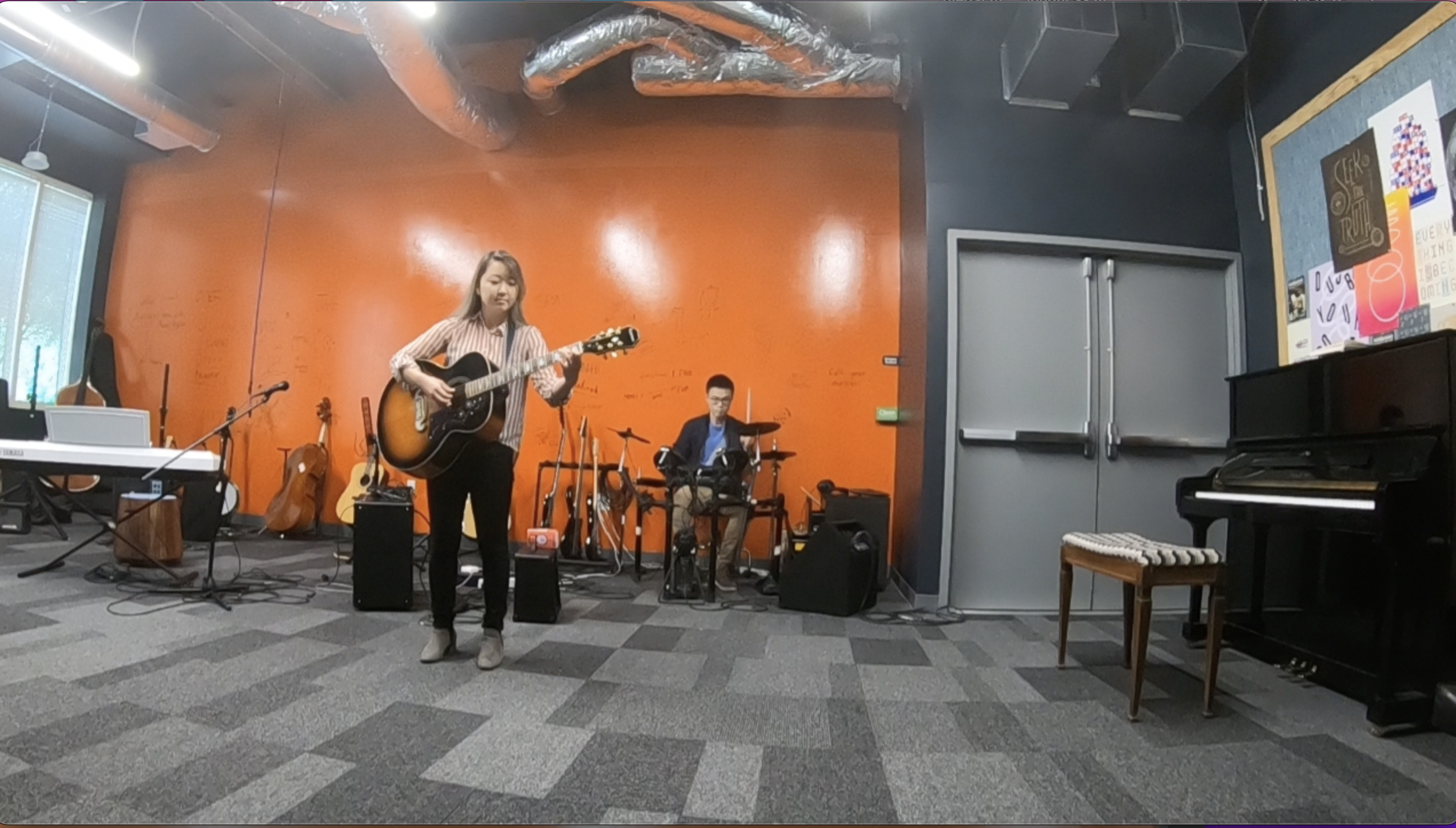} 
        \caption{Raw Image}
        \label{fig:raw_image}
    \end{subfigure}
    \hspace{-0.0085\textwidth} % 控制两张图片间距
    \begin{subfigure}[b]{0.241\textwidth} % 宽度改为 0.4
        \centering
        \includegraphics[width=\textwidth]{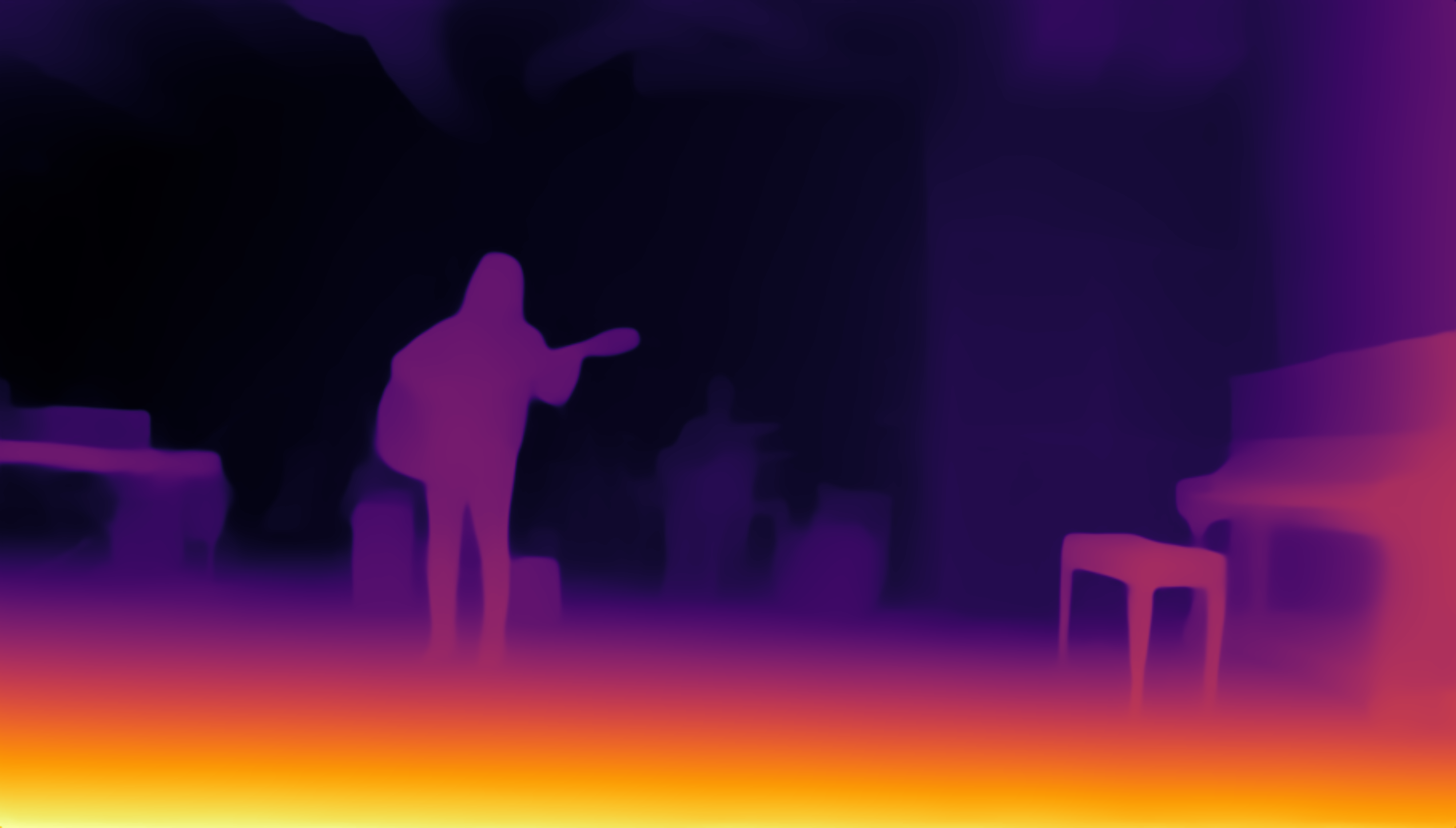} 
        \caption{Depth Anything}
        \label{fig:depth_anything}
    \end{subfigure}
    \caption{Depth estimation model visualization.} 
    
    \label{fig:deep_estimation}
\end{figure}

% \begin{figure}[ht]
%     \centering
%     \includegraphics[width=0.5\textwidth]{deepestimation.png} 
%     \caption{Deep Estimation visualization.}
%     \label{fig:deepestimation}
% \end{figure}

\subsection{Source separation}  \label{sec:sourceseparation}
Due to differences in test datasets, we employ distinct source separation models for various scenarios. For multiple-speaker scenarios, we utilize Conv-TasNet \cite{luo2019conv}, a convolutional time-domain audio separation network renowned for its efficiency and high performance in speech separation tasks. In music scenarios, we adopt Demucs \cite{defossez2019demucs}, which enables the separation of musical tracks such as vocals, drums, and instruments.

\subsection{Audio} \label{sectionaudio}
In the audio processing phase, we handle individual audio tracks that have been separated from multiple audio sources. Utilizing visual cues, we manually align these audio tracks and label them with their corresponding visual elements in the scene, arranged from left to right. This method ensures a coherent integration of audio and visual data, enhancing the perceptual realism of the multimedia experience. The spatial coordinates obtained from the visual component (Sec. \ref{sec:visual}) are utilized to spatialize the audio sources.

To achieve accurate spatialization while maintaining high audio quality, we adopt two methods for audio processing. In the first method, we employ the SADIE II Database\footnote{https://www.york.ac.uk/sadie-project/database.html}} 
 (Subject: KEMAR) \cite{app8112029}, which offers precise Head-Related Transfer Function (HRTF) measurements specifically designed for virtual environments. The spatial coordinates (\textbf{X}, \textbf{Y}, \textbf{Z})  from the visual cues are converted into azimuth and elevation angles. These angles are then used to select the corresponding HRTF files, which are convolved with the individual audio tracks.
 
In the second method, we implement a 3D audio positioning algorithm that enhances the spatial perception of stereo signals by simulating sound propagation in different directions. The algorithm comprises three main steps.\\  
\textbf{Left-Right Positioning}: The algorithm adjusts the signal strength of the left and right channels to control the perceived horizontal position of the sound, effectively creating a stereo panning effect. The signal \(\vec{S}\) represents the input audio as a 1-D mono track, which is split into two channels (left and right) by applying the following ~\eqref{eq:left_right_channel}, where \textbf{X} is a horizontal positioning factor which is calculated from ~\eqref{xy}. 

\begin{equation}
\begin{aligned}
 \text{Left channel} &= x^L = \vec{S} \cdot \frac{1 - \text{\textbf{X}}}{2}, \\
\text{Right channel} &= x^R = \vec{S} \cdot \frac{1 + \text{\textbf{X}}}{2}
\end{aligned}
\label{eq:left_right_channel}
\end{equation}\\
\textbf{Up-Down Positioning}: The elevation is adjusted through frequency filtering, primarily by enhancing or attenuating high frequencies to simulate changes in the sound source's vertical angle. 
 
 In ~\eqref{Up-Down Positioning}, $\mathcal{F}(f)$ represents the frequency adjustment factor, which modifies the energy distribution across frequencies. The \textbf{Y} is the output from the equation (1). The variable $f$ denotes the frequency in Hz. $S(f)$ represents the original frequency spectrum of the stereo signal which is the output from ~\eqref{eq:left_right_channel}, while $S'(f)$ is the adjusted spectrum after applying the frequency adjustment factor.

\begin{equation}
\begin{aligned}
\mathcal{F}(f) &= 1 + \text{\textbf{Y}} \cdot \left(\frac{f}{1000}\right)^{1.5}, \\
S'(f) &= S(f) \cdot \mathcal{F}(f)
\end{aligned}
 \label{Up-Down Positioning}
\end{equation}\\
\textbf{Front-Back Positioning:} 
The algorithm simulates the proximity of the sound source by reducing the volume and adding reverberation. For distant sound sources, the volume is lower, and the reverberation effect is more pronounced. 
\\The output signal \(s_{\text{output}}(t)\) is computed as:
\begin{equation}
s_{\text{output}}(t) = \frac{\text{signal}(t)}{\textbf{Z}} + \alpha \cdot \frac{\text{signal}(t - \Delta t_{\text{samples}})}{\textbf{Z}}
\label{eq:output_signal}
\end{equation}
where
\begin{equation}
\Delta t_{\text{samples}} = \frac{\textbf{Z} \cdot f_s}{v}
\label{eq:time_delay}
\end{equation}
and \(f_s\) is the sampling rate, \(v = 343 \, \text{m/s}\) is the speed of sound, and \(\alpha = 0.3\) is the reverberation intensity factor. \(t\) is the time variable representing the sampling point of the audio signal. \(signal(t)\): The input audio signal at time \(t\). \(signal(t - \Delta t_{\text{samples}})\): A delayed version of the signal.

\section{Experiment}
\subsection{Dataset}  
To evaluate the effectiveness of the system, we utilized two datasets during the testing phase.\\
\textbf{Speech dataset}: The audio stimulus was collected from the LibriSpeech dataset \cite{zen2019libritts} or extracted from multiple-speakers video in YouTube. We designed three scenarios with different speakers from 2 tracks, 3 tracks and 5 tracks.  Each scenario is edited, with its corresponding soundtracks redesigned and synchronized with the visuals. To align the audio with the visual spatial information and provide a reference for evaluation, we manually adjusted the audio panning using the digital audio platform REAPER. This process ensures that the audio accurately corresponds to the spatial positions of visual elements.\\
\textbf{Fair-Play Dataset}: We utilized the FAIR-Play dataset \cite{VisualSound}, which contains 1,871 binaural audio and video clips of musical performances totaling 5.2 hours, for comparative testing in our experiments. 
\subsection{Test process} 
We compare our proposed system with the following baselines: Mono2 Binaural \cite{VisualSound} and Pseudo Binaural  
\cite{xu2021visually}. During the evaluation, each test file was normalized to a loudness level of -23 LUFS, resampled to 16 kHz, and processed using STFT with a Hann window of 25 ms, a hop length of 10 ms, and an FFT size of 512. The audio files were stereo-channel with a bit depth of 16 bits. 

\begin{table*}[ht]
\centering
\caption{Objective evaluation results for speech metrics. The numbers in bold denote the best performance.}
\label{tab:speech_metrics}
\scalebox{1.17}{ 
\begin{tabular}{lcccccccccccc}
\toprule
\multirow{1}{*}{Method} & \multicolumn{3}{c}{MOSNet} & \multicolumn{3}{c}{Nb\_PESQ} & \multicolumn{3}{c}{PESQ} & \multicolumn{3}{c}{STOI} \\
\cmidrule(lr){2-4} \cmidrule(lr){5-7} \cmidrule(lr){8-10} \cmidrule(lr){11-13} 
 Speaker Number 
 & \textbf{2} & \textbf{3} & \textbf{5} 
 & \textbf{2} & \textbf{3} & \textbf{5} 
 & \textbf{2} & \textbf{3} & \textbf{5} 
 & \textbf{2} & \textbf{3} & \textbf{5} \\
\midrule
Mono2Binaural        & 2.814 & 3.447 & 3.022 & 3.706 & 3.634 & 3.078 & 3.423 & 2.215 & 2.089 & 0.976 & 0.912 & 0.898 \\
PseudoBinaural       & \textbf{2.954} & 2.854 & 3.017 & 3.719 & 0.320 & 3.071 & 3.422 & 1.025 & 2.068 & 0.978 & 0.155 & 0.898 \\
Our system (3D) & 2.902 & 3.182 & 2.796 & \textbf{4.019} & 3.136 & 2.957 & \textbf{4.198} & 2.835 & 1.789 & \textbf{0.981} & 0.913 & 0.868 \\
Our system (HRTF)      & 2.797 & \textbf{3.475} & \textbf{3.040} & 3.856 & \textbf{3.718} & \textbf{3.379} & 3.960 & \textbf{2.275} & \textbf{2.558} & 0.979 & \textbf{0.921} & \textbf{0.912} \\
\bottomrule
\end{tabular}}
\end{table*}

\subsection{Evaluation Metrics}

To rigorously assess the performance of our proposed system, we conducted a comprehensive evaluation using multiple metrics that capture both perceptual and objective aspects of audio quality. Specifically, we compared the audio predicted by our system (system-generated) against reference audio samples (collected from the dataset) to determine the fidelity and accuracy of our audio processing algorithm. 
These include: \\ \textbf{STFT Distance:} The Euclidean distance between the complex spectrograms of the reference signal $\mathbf{x}$ and the predicted signal $\tilde{\mathbf{x}}$ for the left and right channels:

\begin{equation}
\mathcal{D}_{\text{STFT}} = \| \mathbf{X}^L - \tilde{\mathbf{X}}^L \|_2 + \| \mathbf{X}^R - \tilde{\mathbf{X}}^R \|_2
\end{equation}
 where $\mathbf{X}^L$, $\mathbf{X}^R$, $\tilde{\mathbf{X}}^L$, and $\tilde{\mathbf{X}}^R$ denotes the complex-valued spectrograms of $\mathbf{x}^L$, $\mathbf{x}^R$, $\tilde{\mathbf{x}}^L$, and $\tilde{\mathbf{x}}^R$, respectively.\\
\textbf{Envelope (ENV) Distance}: Quantifies the Euclidean distance between the envelopes of ground-truth and predicted signals \cite{morgado2018self}. The envelope of the signal \( x(t) \) is represented as \( E[x(t)] \).
\begin{equation}
\begin{split}
\mathcal{D}_{\text{ENV}} = \| E[x^L(t)] - E[\tilde{x}^L(t)] \|_2  
\\ + \| E[x^R(t)] - E[\tilde{x}^R(t)] \|_2 
\end{split} 
\end{equation} \\\textbf{PESQ} (Perceptual Evaluation of Speech Quality): Assesses perceptual audio quality, widely used in telecommunication applications \cite{thiede2000peaq}. \textbf{Nb} stands for Narrowband, referring to a frequency range of 300 Hz to 3400 Hz.\\\textbf{STOI} (Short-Time Objective Intelligibility): Evaluates the intelligibility of speech signals \cite{jensen2016algorithm}.
\\\textbf{MOSNet}: Predicts the Mean Opinion Score (MOS) to estimate perceived audio quality \cite{lo2019mosnet}.

\section{Result and discussion}

Table \ref{tab:speech_metrics} compares the performance of different systems across various speech metrics (MOSNet, NbPESQ, PESQ, and STOI) for 2, 3, and 5 audio tracks. Our proposed system includes two variants: HRTF based and 3D algorithmic approach. While the 3D approach achieves the highest scores in simpler scenarios (e.g., 2Track, with PESQ = 4.198 and NbPESQ = 4.019), the HRTF-based approach demonstrates greater robustness in more complex scenarios. Notably, for the 5Track setup, the HRTF-based approach outperforms the 3D approach in MOSNet (3.040 vs. 2.796) and STOI (0.912 vs. 0.868), highlighting its ability to maintain perceptual quality and intelligibility under challenging multi-speaker conditions. 

\begin{table}[H] 
    \caption{Comparison of STFT and ENV distance metrics for different systems on Speech and Fair-Play datasets. Values in bold represent the best performance.}
    \centering 
    \scalebox{1.2}{ 
    \begin{tabular}{@{}lcccc@{}}
        \toprule
        & \multicolumn{2}{c}{Speech} & \multicolumn{2}{c}{Fair-Play} \\ 
        \cmidrule(lr){2-3} \cmidrule(lr){4-5}
        & STFT & ENV & STFT & ENV \\ 
        \midrule
        Mono2 Binaural & 0.304 & 0.128 & 0.101 & 0.049 \\ 
        Pseudo Binaural & 0.330 & 0.112 & \textbf{0.093} & \textbf{0.048} \\ 
        Our system (3D) &\textbf{0.220} & \textbf{0.094} & 0.151 & 0.063 \\  
        Our system (HRTF) & 2.051 & 0.127 & 0.841 & 0.084 \\ 
        \bottomrule
    \end{tabular}}

    \label{tab:comparison}
\end{table}
Table \ref{tab:comparison} compares the STFT and ENV distance metrics for different systems in the Speech and Fair-Play datasets, highlighting the performance distinctions. The baseline methods, Mono2 Binaural and Pseudo Binaural, achieve consistent results with relatively low STFT and ENV distances, but struggle to capture spatial cues effectively, as indicated by higher STFT values. Pseudo Binaural slightly outperforms Mono2 Binaural in ENV distance on the Fair-Play dataset but exhibits similar limitations on the Speech dataset. In contrast, the proposed system using HRTF achieves competitive ENV distances but the highest STFT values. This is due to the HRTF convolution introducing phase delay, which results in significant phase angle differences and consequently a large STFT distance. Meanwhile, the 3D approach demonstrates superior performance across all metrics and datasets, achieving the lowest STFT and ENV distances.

Furthermore, our system generally achieves lower STFT and ENV distances on the Fair-Play dataset compared to the Speech dataset. This discrepancy can be attributed to the Fair-Play dataset being instrument-based rather than speaker-based, whereas our system relies on object detection in human faces. Additionally, the spatial positioning mismatch between instruments and human faces in the visual component introduces inaccuracies in spatial audio localization, further impacting the performance on the Fair-Play dataset.
 
These results confirm the robustness and effectiveness of our proposed system, particularly the 3D approach, in achieving perceptually accurate and spatially consistent audio synthesis across diverse datasets, underscoring its potential for real-world applications in spatial audio processing. 
\section{Conclusion}
In this paper, we proposed a spatial audio generation system based on visual cues. The system aimed at simplifying the transformation of mono audio into binaural audio in multi-speaker scenarios without relying on binaural dataset. The system integrates object detection, depth estimation, and audio spatialization. Extensive experimental evaluations demonstrate that our system outperforms existing spatial audio generation systems across various metrics. The results highlight significant improvements in spatial consistency between audio and visual components, enhanced speech quality, and robust performance in complex multi-speaker environments.

\bibliographystyle{IEEEtest}
%  \bibliography{icme2025references}
\input{bib.bbl}

\end{document}